\documentclass[aps,prl,twocolumn,letterpaper,superscriptaddress,showpacs]{revtex4-1}
\usepackage{graphicx}
\usepackage{CJK}
\usepackage{verbatim}
\usepackage{physics}
\usepackage[colorlinks,linkcolor=blue,anchorcolor=blue, citecolor=blue,urlcolor=blue,]{hyperref}
\usepackage{lineno} 
\usepackage{bm}
\usepackage{mathptmx}
\usepackage{multirow}
\makeatletter

\begin{document}
\begin{CJK*}{UTF8}{bsmi}
\title{Pressure driven fractionalization of ionic spins results in cuprate-like high-$T_c$ superconductivity in La$_3$Ni$_2$O$_7$}
\author{Ruoshi Jiang (\CJKfamily{gbsn}姜若诗)}
\altaffiliation{These authors contributed equally.}
\affiliation{Tsung-Dao Lee Institute \& School of Physics and Astronomy, Shanghai Jiao Tong University, Shanghai 200240, China}
\author{Jinning Hou (\CJKfamily{gbsn}侯晋宁)}
\altaffiliation{These authors contributed equally.}
\affiliation{Tsung-Dao Lee Institute \& School of Physics and Astronomy, Shanghai Jiao Tong University, Shanghai 200240, China}
\author{Zhiyu Fan (\CJKfamily{gbsn}樊知宇)}
\affiliation{Tsung-Dao Lee Institute \& School of Physics and Astronomy, Shanghai Jiao Tong University, Shanghai 200240, China}
\author{Zi-Jian Lang (\CJKfamily{gbsn}郎子健)}
\affiliation{Tsung-Dao Lee Institute \& School of Physics and Astronomy, Shanghai Jiao Tong University, Shanghai 200240, China}
\author{Wei Ku (\CJKfamily{bsmi}顧威)}
\altaffiliation{corresponding email: weiku@sjtu.edu.cn}
\affiliation{Tsung-Dao Lee Institute \& School of Physics and Astronomy, Shanghai Jiao Tong University, Shanghai 200240, China}
\affiliation{Key Laboratory of Artificial Structures and Quantum Control (Ministry of Education), Shanghai 200240, China}
\affiliation{Shanghai Branch, Hefei National Laboratory, Shanghai 201315, People's Republic of China}

\begin{abstract}
Beyond 14GPa of pressure, bi-layered La$_3$Ni$_2$O$_7$ was recently found to develop strong superconductivity above the liquid nitrogen boiling temperature.
An immediate essential question is the pressure-induced qualitative change of electronic structure that enables the exciting high-temperature superconductivity.
We investigate this timely question via a numerical multi-scale derivation of effective many-body physics.
At the atomic scale, we first clarify that the system has a strong charge transfer nature with itinerant carriers residing mainly in the in-plane oxygen between spin-1 Ni$^{2+}$ ions.
We then elucidate in eV- and sub-eV-scale the key physical effect of the applied pressure: It induces a cuprate-like electronic structure via fractionalizing the Ni ionic spin from 1 to 1/2.
This suggests a high-temperature superconductivity in La$_3$Ni$_2$O$_7$ with microscopic mechanism and ($d$-wave) symmetry similar to that in the cuprates.

\end{abstract}
\maketitle
\end{CJK*}

The families of high-temperature superconductors recently welcome a new member: La$_3$Ni$_2$O$_7$ with a remarkable transition temperature $T_c$ = 80K~\cite{Hualei0516,Hou0719,Yanan0727} under pressure beyond the critical value $P_c\sim14$Gpa. 
As the first material with $T_c$ exceeding the liquid nitrogen boiling temperature since the cuprates, this breakthrough immediately attracts great interest~\cite{Hualei0516,Zhihui0531,Gu0612,Frank0613,Viktor0613,QingGeng0615,Sakakibara0615,Yang0616,Shilenko0626,Zhe0706,Wei0711,Cao0713,Chen0714,Yubo0719,Hou0719,Yang0726,Chen0727,Yanan0727,Yang0728,Hanbit0728,Liao0731,Qu0731,Yang0802,Jiang0813,Zhang0814,Huang0815,Qin0817,Tian0818}.
Specifically, the exciting high-temperature superconductivity is observed in the high pressure phase beyond $P_c$, at which a rather abrupt structural transition takes place (from $Amam$ to $Fmmm$ space group), corresponding to vanishing of tilting of the NiO$_6$ octahedra.
In addition to the emergence of superconductivity, transport properties~\cite{Hualei0516, Hou0719, Yanan0727} also indicates a qualitative change of the electronic structure: from Fermi liquid at low-pressure to strange metal at high-pressure phases.

An immediate essential issue is therefore the pressure-induced qualitative change of electronic structure across the structural transition that enables the exciting high-temperature superconductivity.
As well-known in the field of transition metal oxides~\cite{Anderson1972,Dagotto2005}, a crucial first step is to sort out the local electronic structure that dictates the short-range correlation.
Only under the constraints of such high-energy physics can the rich lower-energy long-range physics emerges, such as the observed strange metal behavior~\cite{Legros2018} or superconductivity~\cite{Dagotto1994}.
Particularly in this case, the qualitative contrast between the low-pressure and high-pressure phases should offer a valuable opportunity to reveal essential ingredients that promote high-temperature superconductivity in general.

Specifically, three key questions deserve most attention at this stage.
First, where do the charge carriers reside, and associated with this what is the effective valence of Ni ion?
Second, what is the main difference in the atomic scale description of the two phases?
Third, how does this difference qualitatively affect the lower-energy physics, particularly concerning the development of high-temperature superconductivity?

Here, we address these timely questions by performing a multi-energy-scale analysis of the local electronic structures in both phases.
At Hartree scale, through realistic one-body spectral function of the \textit{non-collinear} \textit{Curie-paramagnetic} state~\cite{Jiang2022}, we find that this material is in the charge-transfer regime, with self-doped hole carriers residing primarily on the in-plane O atoms.
Within the Ni-O-Ni bi-layers, pressure greatly enhances the inter-layer hopping between Ni-$d_{3z^2-r^2}$ and O-$p_z$ Wannier orbitals~\cite{Wei2002,Marzari1997}.
We then derive the lower-energy effective local Hamiltonians via symmetric canonical transformation~\cite{White2002, Schrieffer1966,Zaanen1988}.
At eV-scale, the above pressure-enhanced hopping produces a exceptionally large inter-layer super-exchange even \textit{exceeding} the intra-atomic exchange.
This in turn fractionalizes the Ni$^{2+}$ ionic spin from 1 to 1/2 at sub-eV scale, leading to an effective Hamiltonian qualitatively identical to the cuprate high-temperature superconductors.
Our result therefore suggests a very similar microscopic superconducting mechanisms and a ($d$-wave) symmetry of the order parameter as in the cuprates.

Let's first consider the Hartree-scale physics of charge distribution, namely the location of the self-doped carriers, and associated with this the dominant valence of Ni ions.
Since such charge physics is of much higher energy than the inter-atomic magnetic correlation, for clarity we evaluate the one-body spectral function in the \textit{Curie-paramagnetic} phase that incorporates carriers' correlation with \textit{unordered} Ni ionic spins via scattering.
This phase is representative of the system at temperature higher than the sub-eV-scale inter-atomic coupling but lower than the eV-scale Hund's coupling.

Specifically, this is performed by ensemble averaging the one-body spectral functions over many \textit{non-collinear} magnetic configurations (typically containing 300 atoms), each with Ni spins pointing toward random directions~\cite{supplementary}.
We used the all-electron implementation~\cite{Blaha2019} of density functional theory~\cite{DFT1, DFT2}, with a typical $U$=6 eV for nickel $d$-orbitals~\cite{Hualei0516, Dudarev1998} in the Hubbard-$U$ functional~\cite{Anisimov1993,Liechtenstein1995}. 
(Our conclusions are insensitive to $U$ within a couple of eV~\cite{Viktor0613}.)
For computational efficiency, we further employ effective interacting Hamiltonian~\cite{supplementary,lang2021,Jiang2022}, $H^{(\mathrm{Hartree})}$, obtained from symmetry-respecting Wannier functions~\cite{Wei2002, Marzari1997}.

\begin{figure}
\centering
\includegraphics[width=0.85\columnwidth]{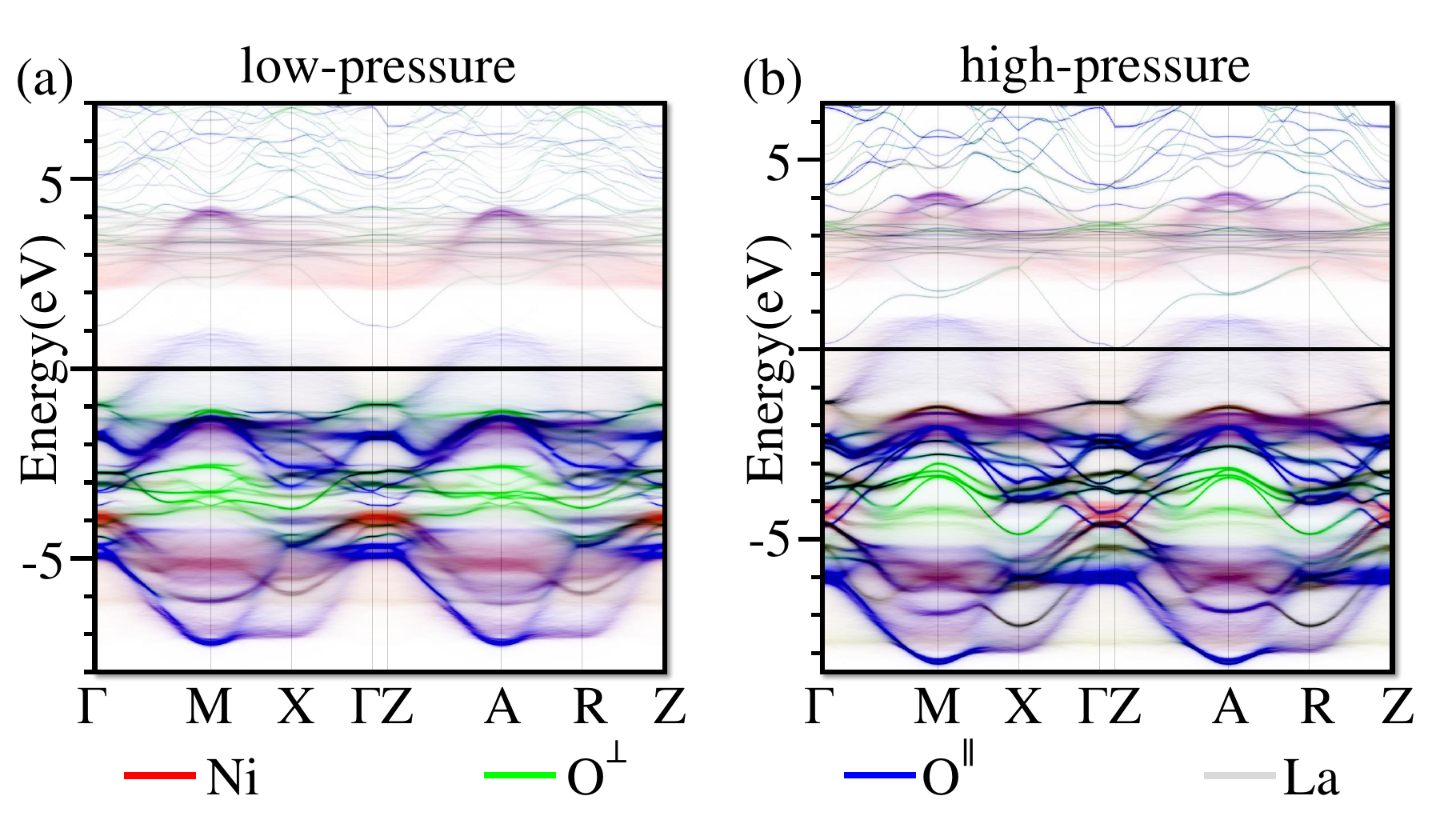}
\vspace{-0.4cm}
\caption{
\textit{Location of itinerant carriers and the importance of correlation with local moments.}
Panel (a) and (b) show the one-body spectral functions of the Curie-paramagnetic state in the low-pressure phase and high-pressure phase, respectively, containing large number of random \textit{non-collinear} Ni-spin directions.
Contributions from orbitals of Ni, La, O$^\perp$ (apical oxygen), and O$^\parallel$ (in-plane oxygen) are represented in different colors.
Notice that only one dominant charge carrier O$^\parallel$ band crosses the Fermi level at zero that strongly scatters against local moments of Ni ions (thus its fuzziness).
}
\vspace{-0.5cm}
\label{fig1}
\end{figure}

Figure~\ref{fig1} shows the resulting one-body spectral function of the low-pressure and high-pressure phases (unfolded~\cite{Wei2010} to the smaller high-pressure unit cell containing two nickel atoms), using the experimental lattice structure at 1.6 and 29.5 GPa~\cite{Hualei0516}.
Notices that only one blue band crosses the chemical potential (set at zero), indicating only \textit{one} dominant type of carriers in the system, primarily composed of O$^\parallel$-$p$ orbitals in the Ni layer.
Also notice the negligible contribution of this band from the apical O$^\perp$ (in green) located between the bi-layer of Ni, indicating that itinerant carriers' charge fluctuations to O$^\perp$~\cite{Frank0613,Shilenko0626,Wei0711,Chen0714} is unessential.
This result unambiguously establishes that the effective charge carriers resides mostly in the in-plane O, as in the cuprate~\cite{Zhang1988} and NdNiO$_2$~\cite{lang2021,lang2022} superconductors.

Consistently, the rather fuzzy red bands in Fig.~\ref{fig1} result from ensemble average of configurations each having 8 occupied and 2 unoccupied Ni-$d$ bands, corresponding to a spin-polarized Ni$^{2+}$-$d^8$ ionic configuration.
The broad spectral distribution of these Ni-$d$ orbitals reflects strong scattering associated with spatial fluctuations of their spin orientation in such a unordered phase.
Furthermore, in Fig.~\ref{fig1} the red Ni-$d$ orbitals have negligible contribution near the chemical potential, indicating their weak charge fluctuation.
Therefore, only their spin and corresponding fluctuation are active at low energy that correlate with the itinerant carriers on O$^\parallel$-$p$.
(Since the inter-ion magnetic coupling is typically beyond 100meV scale for nickelates, Curie-Weiss behavior is not expected below 1000K in experimental magnetic susceptibility~\cite{Zhang2024, wang2023observation}.)

Importantly, correlation with Ni spin still imposes a strong influence on the carrier motion.
For example, notice that in Fig.~\ref{fig1} the quasi-particle peaks of the itinerant carriers are quite broad.
[The broad electron pocket near the $\Gamma$ point are recently observed experimentally~\cite{Yang2023}. As a comparison, observe the sharp quasi-particle peaks for orbitals that correlate weakly with the Ni ions, such as La (in grey) and the apical O at the edge of the bi-layer (in green).]
Consistently, the $\sim2$eV carrier bandwidth in Fig.~\ref{fig1} is also heavily renormalized from the $\sim4$eV bare bandwidth in a non-magnetic~\cite{Hualei0516} or magnetically coherent~\cite{supplementary} phase, consistent with recent experimental observation~\cite{Yang2023}.
Such significant mass enhancement and spectral broadening of quasi-particles reflect the intense scattering of the carriers due to their \textit{strong correlation} with the unordered magnetic moments of Ni ions.

The above analysis indicates that the local bi-layer Ni-O-Ni component is in an effective $\ket{d^8p^6d^8}$ state with itinerant ligand hole residing in the in-plane O.
This conclusion is consistent with some of the previous studies~\cite{Frank0613,Shilenko0626,Chen0714}, upon including charge fluctuation of the ligand holes.
It , however, does not support previous $\ket{d^7d^8}$ (or equivalently $\ket{d^{7.5}d^{7.5}}$)-based theoretical proposals~\cite{Hualei0516,Zhihui0531,Gu0612,QingGeng0615,Sakakibara0615,Yang0616,Yang0726,Hanbit0728,Liao0731,Qu0731,Yang0802,Jiang0813,Zhang0814, Qin0817} or the $\ket{d^6d^9}$-based theory~\cite{Viktor0613}.
The lack of orbital polarization~\cite{Khomskii1973} in experimental structural refinements is in support of our conclusion.

\begin{table}
    \vspace{-0.1cm}
    \caption{
    \textit{Dramatic enhancement of $t^\perp_{pZ}$ at high pressure.}
    Leading in-plane ($\parallel$) and out-of-plane ($\perp$) hopping parameters are given in unit of eV, obtained from DFT extracted Hartree-scale Hamiltonians of the low-pressure and high-pressure phases.
    Subscripts $p$, $X$, and $Z$ denote the O-$p$, Ni-$d_{x^2-y^2}$, and Ni-$d_{3z^2-r^2}$ orbitals, respectively.
    Bold data highlight the dominant pressure induced enhancement.
    }
    \begin{ruledtabular}
        \begin{tabular}{cccccc}
                         & $t^\parallel_{pZ}$ & $t^\parallel_{pX}$ & $t^\parallel_{pp}$ & $t^\perp_{pZ}$       & $t^\perp_{pp}$ \\ \hline
$\mathrm{low\mbox{-}pressure}$  & 0.77       & 1.48              & 0.54              & \textbf{1.39}   & 0.52 \\ 
$\mathrm{high\mbox{-}pressure}$ & 0.91       & 1.62              & 0.55              & \textbf{2.18}   & 0.51 \\
\end{tabular}
    \end{ruledtabular}
    \label{tab1}
    \vspace{-0.4cm}
\end{table}

Now, to address the second question concerning the main difference in the atomic-scale description between the low- and high-pressure phases, Tab.~\ref{tab1} gives the leading hopping parameters obtained from our Wannier function analysis~\cite{supplementary}.
One finds a \textit{dramatic} $\sim60\%$ enhancement of the inter-plane hopping $t^\perp_{pZ}$ between the Ni-$d_{3z^2-r^2}$ and the $p_z$ orbitals of the O$^\perp$ in the middle of the bi-layer structure, from $\sim1.4$eV to $\sim2.2$eV, at least 4 times stronger enhancement than other hopping parameters~\cite{supplementary}.
In great contrast, the most relevant hopping parameter of the itinerant carriers, $t^\parallel_{pX}$ between Ni-$d_{x^2-y^2}$ and in-plane O$^\parallel$-$p_x$/$p_y$ orbitals, only increases by $\sim10\%$ at high pressure.

Such a large enhancement is likely closely associated with the qualitative change of the low-energy electronic structure at high pressure.
A straightforward scenario is the formation of strong bonding of Ni-$d_{3z^2-r^2}$ orbitals across layers~\cite{Hualei0516,Gu0612,Sakakibara0615,Yang0616,Yang0726,Yang0728,Hanbit0728,Liao0731,Zhang0814} via a strong $t^\perp_{pZ}$ that overcomes the charge transfer potential energy $E_\mathrm{CT}=E(\ket{d^9 p^5 d^8})-E(\ket{d^8 p^6 d^8})$.
However, Fig.~\ref{fig1} shows clearly that $E_\mathrm{CT}$, roughly the energy difference between the red unoccupied Ni-$d_{3z^2-r^2}$ orbital and the green occupied O$^\perp$-$p_z$ orbital, are about 3-4eV.
So, even in the high-pressure phase this system is far from the $t^\perp_{pZ} \gg E_\mathrm{CT}$ bonding limit~\cite{Zhihui0531,Qu0731}.
Clearly, if even the largest kinetic coupling of the system $t^\perp_{pZ}$ between the \textit{neighboring} Ni and O atoms is unable to drive a purely bonding picture, strong bonding \textit{across bi-layer} Ni atoms surely cannot be realistic.
One thus has to seek a more decisive mechanism in lower-energy scale to reveal the physical effect of pressure.

Before digging deeper into the lower-energy physics, first notice that the above enhancement of the bi-layer Ni-O-Ni coupling does not directly contribute to carriers' motion.
The former concerns the Ni-$d_{3z^2-r^2}$ and O$^\perp$-$p_z$ orbitals across layers, while the latter mostly the O$^\parallel$-$p_x$/$p_y$ and Ni-$d_{x^2-y^2}$ orbitals in the same layer.
The kinetic processes between these two sub-spaces are heavily suppressed not only because of the weaker hopping over energy difference between them, but also due to carriers' $x^2-y^2$ symmetry around each Ni (due to strong coupling to Ni-$d_{x^2-y^2}$ orbitals.)
As discussed above, the lack of green contribution in the itinerant band in Fig.~\ref{fig1} clearly indicates such decoupling in kinetics.

Therefore, the primary effects of the above $t^\perp_{pZ}$ enhancement on lower-energy physics should be through the qualitative change in the local electronic structure of bi-layer Ni-O-Ni component and its correlation with the itinerant carriers.
Indeed, while in these two phases Fig.~\ref{fig1} shows very similar dispersion of the carrier bands crossing the chemical potential, distinct broadening occurs especially along the X-R $(\pi,0,k_z)$ path, implying different correlation strength with the local electronic structure of the bi-layer Ni-O-Ni component.

We thus proceed to investigate the \textit{local} many-body electronic structure of the bi-layer Ni-O-Ni component, aiming at identifying the bifurcation in low-energy behavior that qualitatively distinguishes the low- and high-pressure phases~\cite{note_hole_impact}.
We start with above $H^{(\mathrm{Hartree})}$ in atomic Wannier basis~\cite{Wei2002,supplementary} containing intra-atomic Coulomb interaction.
We then derive its eV-scale effective Hamiltonian within the local $\ket{d^8p^6d^8}$ subspace, by numerically `integrating out' states containing Ni-$d^9$ configurations.

Specifically, this is executed by numerically applying a series of \textit{symmetric} unitary transformations to completely zero out the coupling between the high-energy states from the remaining low-energy subspace~\cite{supplementary}.
Other than taking special care on maintaining explicitly the symmetry of the representation, this numerical procedure is a straightforward implementation of the numerical canonical transformation~\cite{White2002}, which itself can be considered a complete \textit{non-perturbative} extension of the Schrieffer-Wolff transformation~\cite{Schrieffer1966,Zaanen1988,lang2021,Yin2009}.
The completeness of the resulting effective description is easily checked by comparing its eigenstates against the low-energy ones of the original Hamiltonian.

Given the fully occupied O$^\perp$-$p$ shell upon integrating out states containing Ni-$d^9$ configurations, among the local $\ket{d^8p^6d^8}$ states the remaining low-energy subspace naturally has no charge freedom left.
The resulting eV-scale local effective Hamiltonian is therefore spin-only and dominated by Heisenberg couplings between four electrons that each singly occupy an $e_g$ orbital of the Ni-$d$ shell:
\begin{equation}
        H^{\mathrm{(eV)}} = -\Tilde{J}_\mathrm{H}~\mathbf{S}_X^{(1)}\cdot\mathbf{S}_Z^{(1)}\\
        + J_{ZZ}~\mathbf{S}_Z^{(1)}\cdot\mathbf{S}_Z^{(2)}\\
        - \Tilde{J}_\mathrm{H}~\mathbf{S}_Z^{(2)}\cdot\mathbf{S}_X^{(2)},
    \label{eq_4spin}
\end{equation}
where in both Ni ions [labeled by (1) and (2)] the ferromagnetic Hund's coupling between spins in the singly occupied $d_{x^2-y^2}$ orbital, $\mathbf{S}_X$, and $d_{3z^2-r^2}$ orbital, $\mathbf{S}_Z$, in the same Ni-$d$ shell is renormalized to a weaker $\Tilde{J}_\mathrm{H}$ by the higher-energy charge fluctuation.
In addition, an effective anti-ferromagnetic coupling, $J_{ZZ}$, emerges between the Ni ions via the super-exchange mechanism~\cite{Anderson1950}.

\begin{table}
    \vspace{-0.3cm}
    \caption{
    \textit{Pressure-induced fractionalization of Ni$^{2+}$ ionic spin.}
    Magnetic coupling strengths in emerged $H^{\mathrm{(eV)}}$ and $H^{\mathrm{(sub\text{-}eV)}}$ are given in unit of eV for the low-pressure and high-pressure phases, derived from Hartree-scale Hamiltonian~\cite{supplementary} with parameters $E_\mathrm{CT}\sim 3$, $U_{d}=6$, $U_{p}=4$~\cite{Ogata_2008} and the bare $J_{\mathrm{H}}=1$.
    Notice the pressure-induced fractionalization of Ni$^{2+}$ effective ionic spin from (bold) 1 to 1/2.}
    \begin{ruledtabular}
        \begin{tabular}{c @{\hspace{0.5cm}} |ccc @{\hspace{0.5cm}}|ccc}
          & \multicolumn{3}{c|}{$H^{\mathrm{(eV)}}$} &\multicolumn{3}{c}{$H^{\mathrm{(sub\text{-}eV)}}$}\\
         &$\tilde{J}_{\mathrm{H}}$ & & $J_{ZZ}$ &  $S_{\mathrm{eff}}$  & & $J^{\perp}$   \\ 
        \hline
$\mathrm{low\mbox{-}pressure}$   & 0.86 & $>$ & 0.30 & \textbf{1} & & 0.08  \\ 
$\mathrm{high\mbox{-}pressure}$  & 0.79 & \textbf{$<$} & 0.96 & \textbf{1/2}  & & 0.09  \\
        \end{tabular}
    \end{ruledtabular}
    \label{tab2}
    \vspace{-0.4cm}
\end{table}

Interestingly, Tab.~\ref{tab2} shows an important distinction in $H^{(\mathrm{eV})}$ between the low- and high-pressure phases.
At low-pressure, even though a rather strong anti-ferromagnetic inter-atomic super-exchange coupling, $J_{ZZ}\sim0.3$eV, emerges between the $d_{3z^2-r^2}$ orbitals, it is \textit{as usual} in no competition with the ferromagnetic intra-atomic Hund's coupling $\Tilde{J}_\mathrm{H}\sim 0.9$eV, which is only slightly screened from its bare value (set to be 1eV for simplicity).
In great contrast, at high-pressure, the enhanced $t^\perp_{pZ}$ leads to a \textit{highly unusual} situation, in which the \textit{inter-layer} super-exchange $J_{ZZ}\sim1$eV is even stronger than the intra-atomic Hund's coupling, $\tilde{J}_\text{H}\sim0.8$eV.
(Inclusion of itinerant carriers' additional fluctuation to the $d_{x^2-y^2}$ orbital can only further weaken $\tilde{J}_\mathrm{H}$ and in turn strengthen the high-pressure behavior~\cite{supplementary}.)
Such switching of dominant couplings is the direct indication we seek for a qualitative change of low-energy behavior!

To more explicitly demonstrate the qualitatively distinct physical behavior, let's further integrate out the leading local physics in both phases.
At low pressure, the dominant Hund's coupling points to the intra-atomic singlet as the high-energy states to be integrated out.
After applying the above numerical canonical transformation again~\cite{supplementary}, only states with the spin-1 triplets, $\mathbf{S}_{\mathrm{eff}}$ in each Ni ion survive at low-energy, described by a sub-eV-scale local effective Hamiltonian,
\begin{equation}
        H^{\mathrm{(sub\text{-}eV)}} = J^{\perp}~\mathbf{S}_{\mathrm{eff}}^{(1)}\cdot\mathbf{S}_{\mathrm{eff}}^{(2)},
    \label{eq_2spin}
\end{equation}
with $J^{\perp}\sim0.1$eV as shown in Tab.~\ref{tab2}.
In other words in sub-eV scale, the local Ni-O-Ni component for the low-pressure phase resembles the typical Ni$^{2+}$ nickelates\cite{Greenwood1984}, consisting of spin-1 Ni ions that couple to each other via 100meV scale anti-ferromagnetic super-exchange.

With such a large coupling between spin-1 ions, one expects a strong inter-layer anti-ferromagnetic correlation, or even a long-range magnetic order in the low-pressure phase.
Typically, such magnetic order seriously suffers from carrier-induced long-range quantum fluctuation~\cite{Tam2015} and is not expected to survive~\cite{Tan2022,Hou2023} the significant self-doping (1 hole per 2 Ni atoms) in the material.
However, since the normal-state Fermi surface of the carrier band in Fig.~\ref{fig1} turns out to be very well nested via $\Delta \mathbf{k}\sim(\pi,\pi, 0)$~\cite{supplementary}, the common G-type anti-ferromagnetic order may still be a stable phase with dramatically reduced carrier density.
Indeed, current magnetic susceptibility~\cite{Liu2022} and resistivity~\cite{Liu2022,Hualei0516,Zhe0706,Yanan0727} measurements found a ``kink'' at around $T=150$K, consistent with the emergence of a G-type anti-ferromagnetic order.
Similarly, recent optical conductivity~\cite{Zhe0706} indeed found a surprisingly low carrier density as expected here.

In great contrast, for the high-pressure phase, the dominant inter-layer anti-ferromagnetic super-exchange $J_{ZZ}$ pushes the triplet states of the two $\mathbf{S}_Z$ to high energy, leaving only a robust singlet between them at low energy.
Consequently, while the resulting sub-eV-scale local Hamiltonian takes the same form as Eq.~\ref{eq_2spin}, $\mathbf{S}_{\mathrm{eff}}$ is instead only 1/2 from the remaining $\mathbf{S}_X$ in each Ni ion.
In other words, qualitatively distinct from that in the low-pressure phase, in sub-eV scale, the Ni$^{2+}$ ionic spin is `fractionalized' from 1 to 1/2, analogous to the fractionalization of spin-1 local moments in the AKLT models~\cite{Affleck1987} or spins and carriers in some other strongly correlated systems~\cite{FADDEEV1981,Laughlin1983,Kivelson1987,KITAEV2006}.

\begin{figure}
\centering
\vspace{-0.6cm}
\includegraphics[width=0.8\columnwidth]{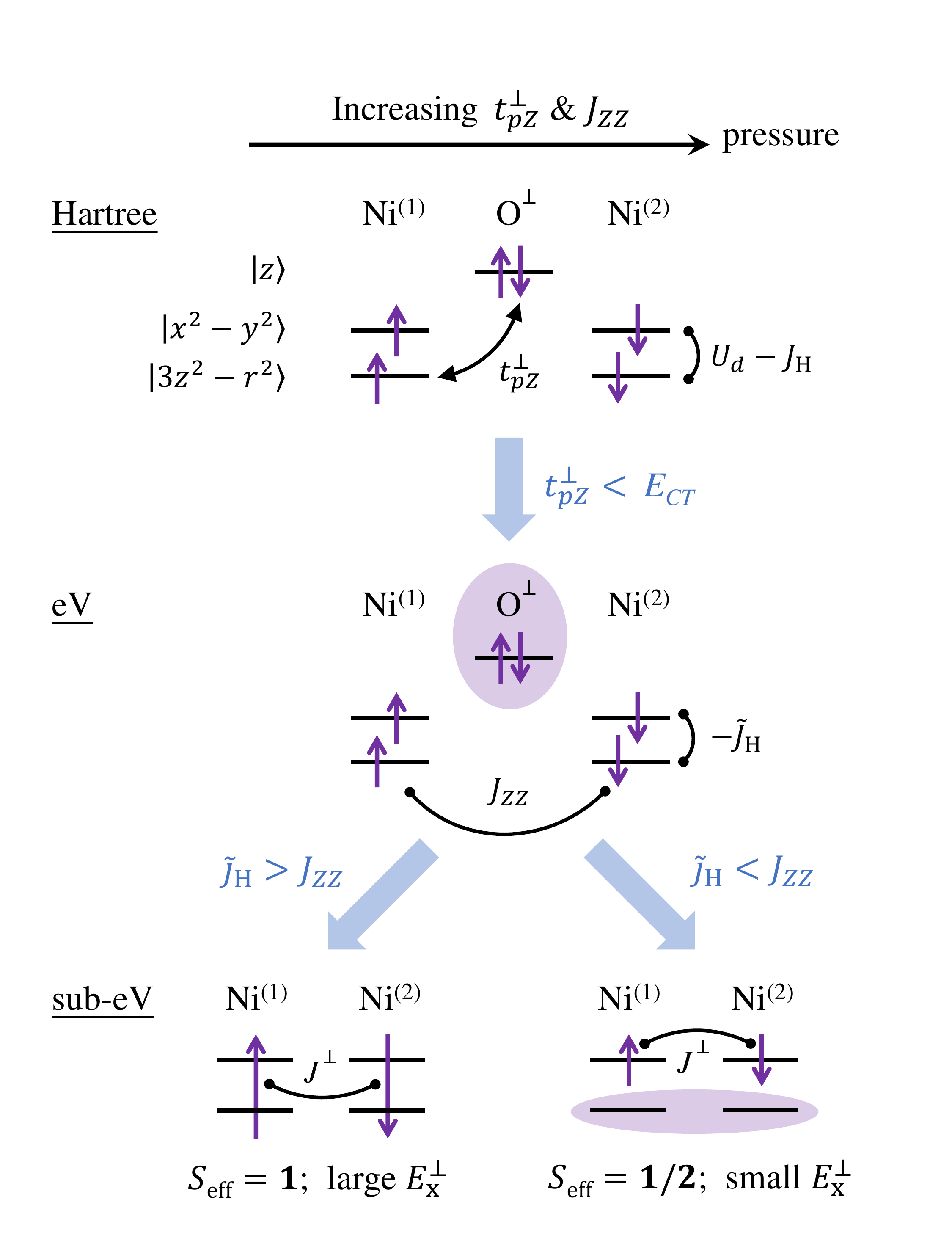}
\vspace{-0.4cm}
\caption{\textit{Illustration of the multi-scale emergence of pressure-induced fractionalization of Ni$^{2+}$ ionic spin in the local bi-layer Ni-O-Ni component.}
At Hartree scale, the dominant energy is the charge-transfer energy to Ni-$d^9$ configuration.
At eV scale, the charge freedom of Ni-$d^9$ is therefore frozen and the fully occupied O$^\perp$ orbital becomes inactive (shadowed), leaving only spins dynamics of Ni orbitals active.
At low-pressure, the stronger screened Hund's coupling $\Tilde{J}_\mathrm{H}$ dictates a spin-1 Ni ions at sub-eV scale with large inter-layer exchange energy $E_\mathrm{x}^\perp\sim 3 J^\perp \sim 250meV$.
In contrast, at high-pressure, an unusually strong inter-layer super-exchange $J_{ZZ}$ emerges and causes an inactive (shadowed) spin-0 singlet of $d_{3z^2-r^2}$ orbitals.
The sub-eV description therefore resembles that of the high-$T_\mathrm{c}$ cuprates, with a `fractionalized' effective 1/2 spin of Ni$^{2+}$ ions.
}
\vspace{-0.5cm}
\label{fig2}
\end{figure}

We thus have successfully identified the key physical impact of the applied pressure that leads to qualitative distinct low-energy behaviors in the local bi-layer Ni-O-Ni component.
As illustrated in Fig.~\ref{fig2}, in the low-pressure phase, Ni$^{2+}$ ions are in the usual spin-1 $d^8$-configuration dictated by Hund's coupling.
In the high-pressure phase, on the other hand, the dramatically enhanced $t^\perp_{pZ}$ promotes a much stronger $J_{ZZ}$ of eV-scale, which in turn drives the formation of a singlet across the Ni ions.
In the sub-eV scale, this effectively removes half of the spin degree of freedom of the Ni$^{2+}$ ions (without dramatically altering their charge) and significantly reduces the emerged super-exchange energy across layers.

Such fractionalization the Ni$^{2+}$ ionic spin from 1 to 1/2 in the high-pressure phase leads to an extremely interesting consequence.
That is, the sub-eV-scale electronic Hamiltonian of the system now \textit{exactly} resembles that of the hole-doped high-temperature superconducting cuprates in a `decorated square lattice', with ligand holes as carriers residing between sites and magnetically correlating with the spin-1/2 transitional metal ions on the sites~\cite{Zaanen1985,Zhang1988,Emery1988}.
Given such a strong resemblance in the sub-eV-scale Hamiltonian, the observed emergence of high-temperature superconductivity in the high-pressure phase of La$_3$Ni$_2$O$_7$ seems more of a reasonable outcome than an unexpected surprise.
In fact, it can be viewed as realization of the previous proposal~\cite{lang2022} to improve the superconducting properties of spin-1/2 nickelates Nd$_{0.8}$Sr$_{0.2}$NiO$_2$~\cite{Li2019} by enhancing its charge-transfer nature. 
Consistently, the observation of a strange metal behavior~\cite{Hualei0516,Hou0719,Yanan0727} (temperature $T$-linear resistivity) in the high-pressure phase seems natural as well, since the same is also observed in the hole-doped cuprates~\cite{Legros2018,Daou2008,Cooper2009, Tao2021}.
Our study therefore suggests a very similar superconductivity to the cuprates, including its microscopic mechanisms and a ($d$-wave) symmetry of the order parameter~\cite{Tsuei1994}.

In comparison, the robust local magnetic structure in the low-pressure phase, with larger local moments and stronger couplings, resembles more the typical doped nickelates with $d^8$ configuration, such as Eu$_{0.9}$Sr$_{1.1}$NiO$_4$~\cite{Uchida2011}, Nd$_{0.8}$Sr$_{0.2}$NiO$_3$~\cite{Li2019}, that do not exhibit superconductivity.
The lack of superconductivity in the low-pressure phase of La$_3$Ni$_2$O$_7$ therefore seems quite natural as well.
Clearly, the contrast in the sub-eV electronic structure between the high- and low-pressure phases provides valuable clues about the essential ingredients of unconventional superconductivity.

Note that our analysis also offers a microscopic mechanism for the observed abrupt structural transition at $\sim14$Gpa.
In the low-pressure phase, since the dominant local physics at the eV scale is the \textit{intra-atomic} Hund's coupling (c.f. Eq.~\ref{eq_4spin}), the system naturally finds the tilted octahedra to optimize the potential energy by reducing the empty space in the structure.
On the other hand, in the high-pressure phase, the dominant local physics at the eV scale becomes the \textit{inter-layer} super-exchange, which allows a significant magnetic energy gain ($\sim\frac{3}{4}J_{ZZ}$~\cite{AshcroftMermin} per Ni-O-Ni component) through formation of the singlet.
The high-pressure phase therefore prefers maximizing $J_{ZZ}$ through enhancing $t^\perp_{pZ}$ by straightening the Ni-O-Ni bonds.
Given the qualitatively distinct eV-scale electronic structure in these two phases, an abrupt first-order-like phase transition is thus expected, in agreement with experimental observation~\cite{Hualei0516, Hou0719}.

In summary, to reveal the main physical effect of applied pressure on La$_3$Ni$_2$O$_7$ that promotes the recently discovered high-temperature superconductivity, we perform a multi-energy-scale study of the many-body electronic structure.
At Hartree-scale, we first clarify the charge transfer nature of the system, with self-doped hole residing in the in-plane O between Ni$^{2+}$ ions in $d^8$ configuration.
We then elucidate in eV- and sub-eV-scale the key physical effects of the applied pressure: driving a cuprate-like electronic structure through fractionalizing the Ni ionic spin from 1 to 1/2 via inter-layer super-exchange.
Our results therefore suggest a high-temperature superconductivity in La$_3$Ni$_2$O$_7$ with a similar microscopic mechanism and a ($d$-wave) symmetry as the cuprates.

This work is supported by National Natural Science Foundation of China (NSFC) under Grant Nos. 12274287 and 12042507, and Innovation Program for Quantum Science and Technology No. 2021ZD0301900.

\vspace{0.5cm}
\begin{center}
\textbf{\large Supplemental materials}
\end{center}
\setcounter{equation}{0}
\setcounter{figure}{0}
\setcounter{table}{0}
\makeatletter
\renewcommand{\theequation}{S\arabic{equation}}
\renewcommand{\thetable}{S\arabic{table}}
\renewcommand{\thefigure}{S\arabic{figure}}

\subsection{1. The comparison of the crystal structure under pressure}

According to Ref.~\cite{Hualei0516}, the crystal structures of the low-pressure and high-pressure phase corresponds to space group $Amam$ and $Fmmm$, respectively.
As shown below, the main difference is the straightening of the Ni-O-Ni bond within the bi-layer structure.

\begin{figure}[h]
	\begin{center}
	\includegraphics[width=0.8\columnwidth]{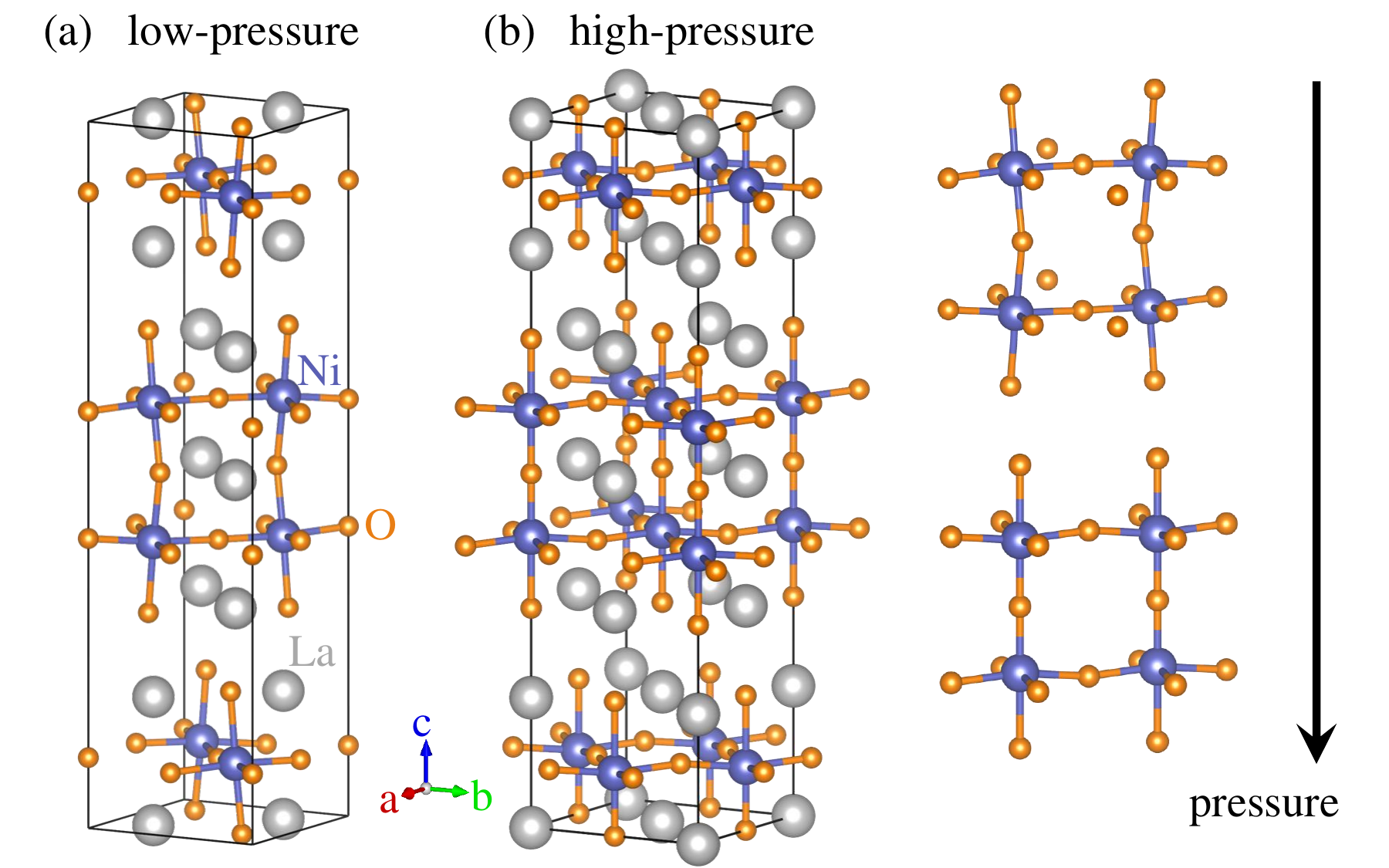}
	\end{center}
	\vspace{-0.5cm}
	\caption{
    Comparison of crystal structures of the low-pressure and high-pressure phases.
	}
	\label{figs0}
	\vspace{-0.3cm}
\end{figure}

\subsection{2. Computational details of density functional calculation}

For this prototypical case, we obtain the most relevant Hilbert space within $15$ eV from the spin polarized LDA+$U$ ~\cite{Anisimov1993, Liechtenstein1995} electronic structure of the bi-layer La$_3$Ni$_2$O$_7$, using the linearized augmented plane wave~\cite{Singh} implementation~\cite{Blaha1990} of the density functional theory (DFT)~\cite{DFT1, DFT2}.
We take from Ref.~\cite{Hualei0516} the experimental lattice structure with lattice constant $a=5.4392\mathring{A}, b=5.3768\mathring{A}, c=20.403\mathring{A}$ within the space group $Amam$ in low pressure 1.6GPa, and the lattice constant $a=5.289\mathring{A}, b=5.218\mathring{A}, c=19.734\mathring{A}$ within the space group $Fmmm$ in high pressure 29.5GPa.
Following the previous study~\cite{Hualei0516, Dudarev1998}, we pick a typical $U-J$ = 6 eV for Ni $d$-orbitals.
We checked that our qualitative results are insensitive to this parameter up to at least 1eV of variation.

\subsection{3. Computing band structure in the non-collinear Curie-paramagnetic phase}

For transition metal oxides with open $d$-shell, the standard local density approximation (LDA) suffers from unphysically strong charge fluctuation of these $d$-orbitals.
The LDA+$U$ offers an inexpensive alternative to encode a necessary energy scale of charging and generate more reasonable total density.
However, the typical implementation of LDA+$U$ requires the transition metal ion to be spin-polarized, forcing the results to unnecessarily adhere to an particular long-range order.

This standard operation becomes problematic when the actual system has fluctuating local moments without long-range magnetic order, such as in a quantum Curie-paramagnetic phase of strongly correlated metals or geometrically frustrated magnets.
Or, this operation would mask the physics of interest, for example the charge carrier distribution, when the physics is of much higher energy scale than inter-atomic magnetic correlation and should be insensitive to a particular long-range magnetic order.
In these cases, it is highly beneficial to have access to the band structure through the one-body spectral function in a simulated (quantum) Curie-paramagnetic phase, in which inter-atomic magnetic correlation is negligible, such that impacts of the (artificial) magnetic order is removed.

\begin{figure}[h]
	\begin{center}
    \vspace{-0.5cm}
	\includegraphics[width=0.85\columnwidth]{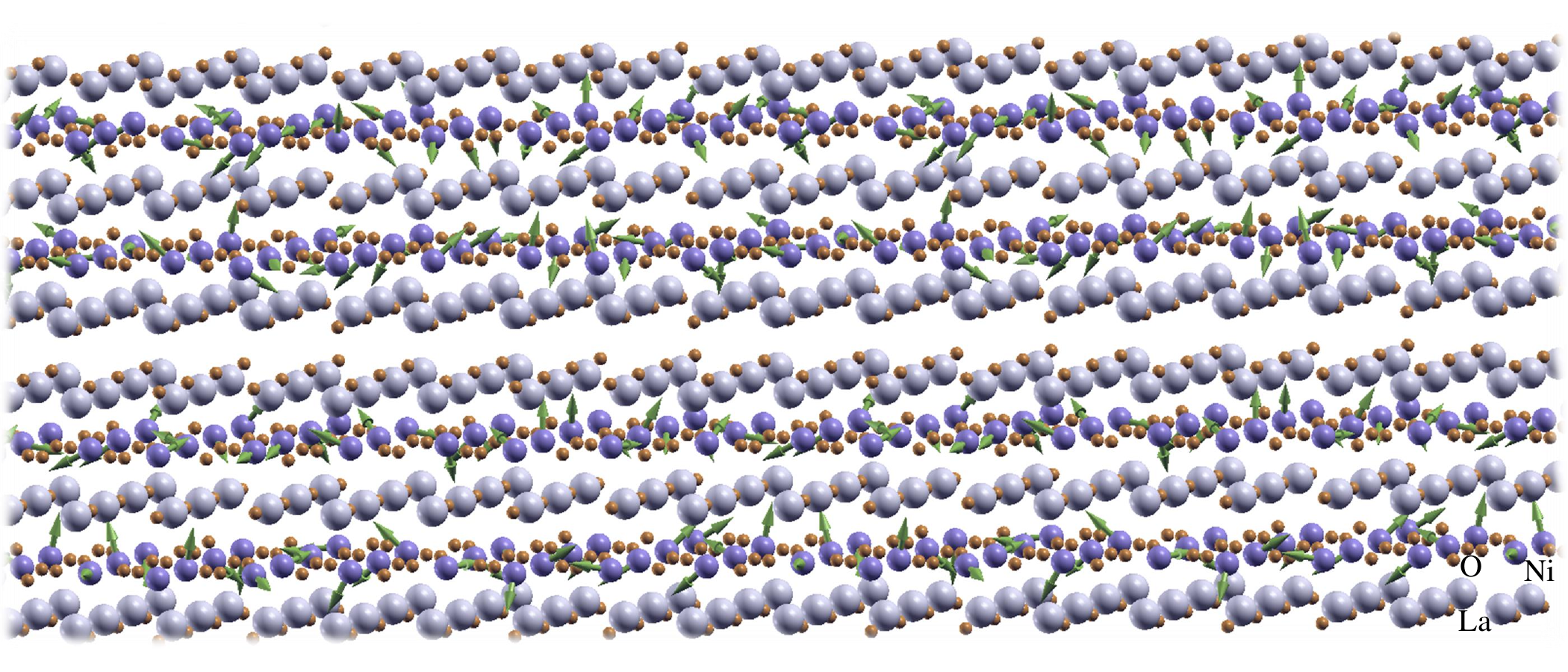}
	\end{center}
	\vspace{-0.5cm}
	\caption{
    Illustration of \textit{unordered non-collinear} Curie paramagnetic configurations of low-pressure phase with spin directions of each magnetic atom Ni denoted by the green arrows.
	}
	\label{figs3}
	\vspace{-0.3cm}
\end{figure}

This one-body spectral function can be obtained via standard DFT packages, by averaging the `unfolded' one-particle spectral function~\cite{Wei2010} of an ensemble of large super cells, each containing disordered \textit{non-collinear} spin orientation of each transition metal ions.
Naturally, the smallness of magnetic correlation of each super cell should be checked via magnetic correlation function $\chi(r) = \sum_i\mathbf{S}_{i+r} \cdot \mathbf{S}_i$.
Figure~\ref{figs3} gives an example of such super cell that is actually used in this study.
In addition, it helps to reduce the artificial periodicity of the super cells if they are chosen in different orientation.

Additionally, one can greatly simplify the above procedure if a SU(2)-symmetric interacting Hamiltonian is obtained first, via for example the method described in the subsession 5 below.
While this is physically not necessary, the reduced Hilbert space does reduce dramatically the computation expense, allowing even larger supercells that resemble better the magnetically disordered systems.


\subsection{4. Well-nested Fermi surface in low-pressure phase}

\begin{figure}[h]
	\begin{center}
	\includegraphics[width=0.65\columnwidth]{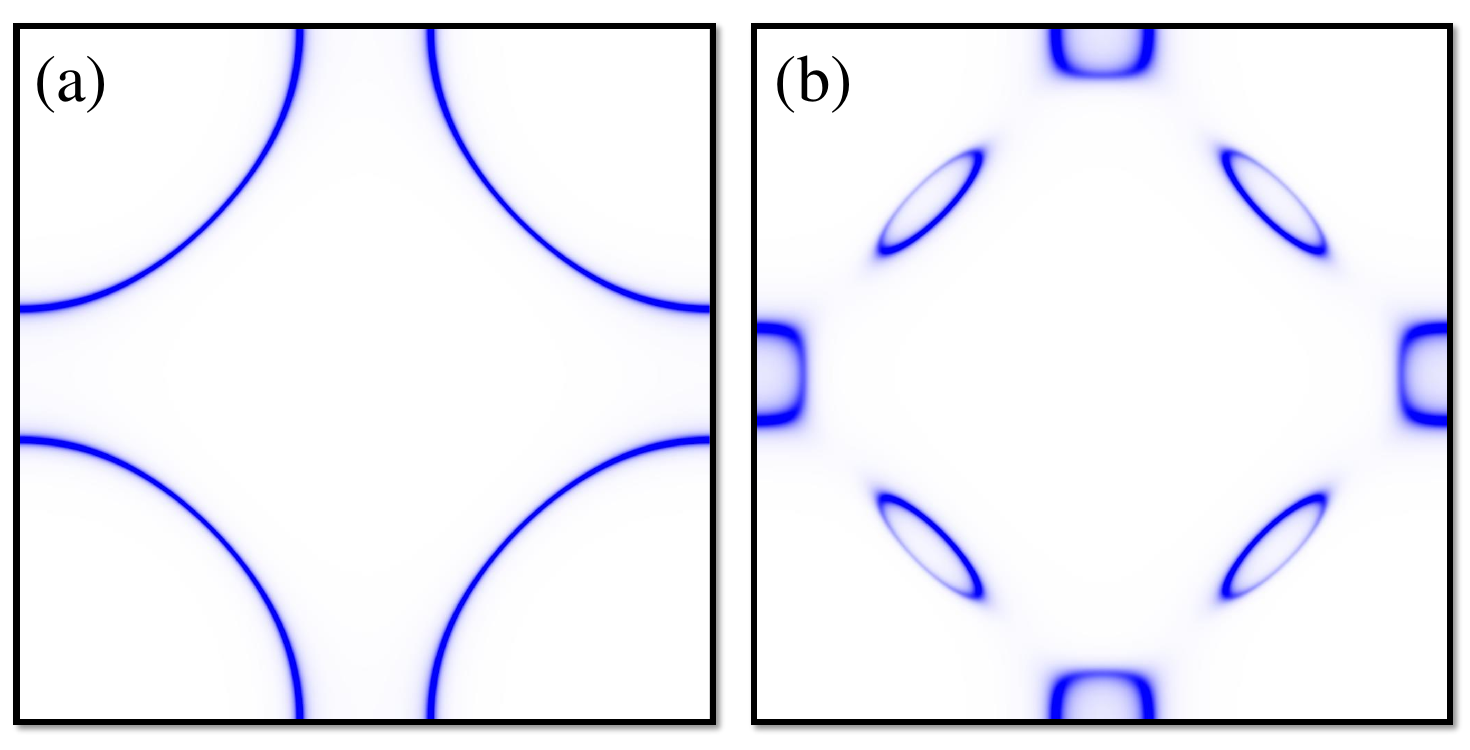}
	\end{center}
	\vspace{-0.4cm}
	\caption{Illustration of (a) the well-nested Fermi surface of the itinerant ligand hole carriers in the Curie-paramagnetic phase at low pressure and (b) the emerged small electron and hole Fermi pockets upon coupling to a G-type anti-ferromagnetic order. This indicates a significant reduction of carrier density in the magnetic phase, and associated with it a weak itinerant carrier-induced long-range magnetic fluctuation. The magnetic order of the spin-1 Ni$^{2+}$ ions is therefore more stable with such well-nested Fermi surface.
	}
	\label{figs4}
	\vspace{-0.5cm}
\end{figure}

In Fig.1, the itinerant band in the Curie-paramagnetic phase has a rather well-nested Fermi surface.
Fig.~\ref{figs4} illustrates the corresponding Fermi surface obtained from the main dispersion in Fig.1, map-able to renormalized hopping parameters $(t_1, t_2, t_3, t_z)\sim(-0.225, 0.01, -0.03, \sim 0)$eV.
Upon coupling to a $(\pi,\pi,\pi)$ G-type magnetic order, the Fermi surface is strongly gapped out, leaving only small electron- and hole-Fermi pockets.
One thus expects a significantly reduced carrier density in the ordered phase (or its proximity).
Related to this, the carrier-induced long-range quantum fluctuation of the magnetic order is strongly suppressed, allowing the long-range order to potentially survive even with 0.5/Ni self-doing of this material.

\subsection{5. Atomic Wannier orbitals as basis for $H^{\text{(Hartree)}}$}

The Hartree-scale physics is most sensible using atomically local orbitals as basis.
We therefore construct a complete set of atomic-like Ni $d$-, O $p$- and La $d$-, $f$-Wannier orbitals~\cite{Wei2002, Wei2006, Marzari1997} without down-folding to one-body low-energy subspace.
These symmetry-respecting Wannier orbitals form a nearly complete basis that are atomically local and nearly configuration independent, making them ideal for the ensemble averaging of resulting one-particle spectral function.

\subsection{6. Extraction of interacting $H^{\text{(Hartree)}}$ from DFT results}

It is well known that for transition metal oxides, the open-shell $d$-electrons have heavily suppressed charge freedom due to the large intra-atomic repulsion $U$.
It is often necessary to apply \textit{spin-polarized} DFT+$U$ method to remove most of them from the Fermi surface in order to obtain a good density within DFT.
The spin-polarization of the atoms however does not provide a straightforward route to a generic SU(2) \textit{symmetric} interacting $H^{\text{(Hartree)}}$:
\begin{equation}
    \begin{split}
        H^{(\mathrm{Hartree})} &= \sum_{i,i^\prime,m,m^\prime,\nu}t_{ii^\prime mm ^\prime}c^\dagger_{im\nu}c_{i^\prime m^\prime \nu}\\
        &+\frac{1}{2}\sum_{i,m,m^\prime,m^{\prime\prime},m^{\prime\prime\prime},\nu,\nu^\prime} U_{m m^{\prime\prime}  m^\prime m^{\prime\prime\prime}} c^\dagger_{im\nu}c^\dagger_{im^{\prime\prime}\nu^\prime}c_{im^{\prime\prime\prime}\nu^\prime}c_{im^\prime\nu}\\
        &-\sum_{i,m,m^\prime,\nu} V^\text{HF0}_{m m^\prime} c^\dagger_{im\nu}c_{im^\prime\nu},
    \end{split}
    \label{eq1}
\end{equation}
where $t$ denotes the one-body hopping strength,
$U$ the intra-atomic two-body Coulomb interaction of Ni, and $c^\dagger_{im\nu}$ the creation operators of orbitals $m$ and spin $\nu$ at lattice site $i$.
Note that for convenience, $t$ also includes the additional contribution of spin-averaged Hartree-Fock meanfield, $V^\text{HF0}$, so the latter is subtracted in the last line of Eq.~\ref{eq1}.

\begin{figure}[h]
	\begin{center}
	\includegraphics[width=1\columnwidth]{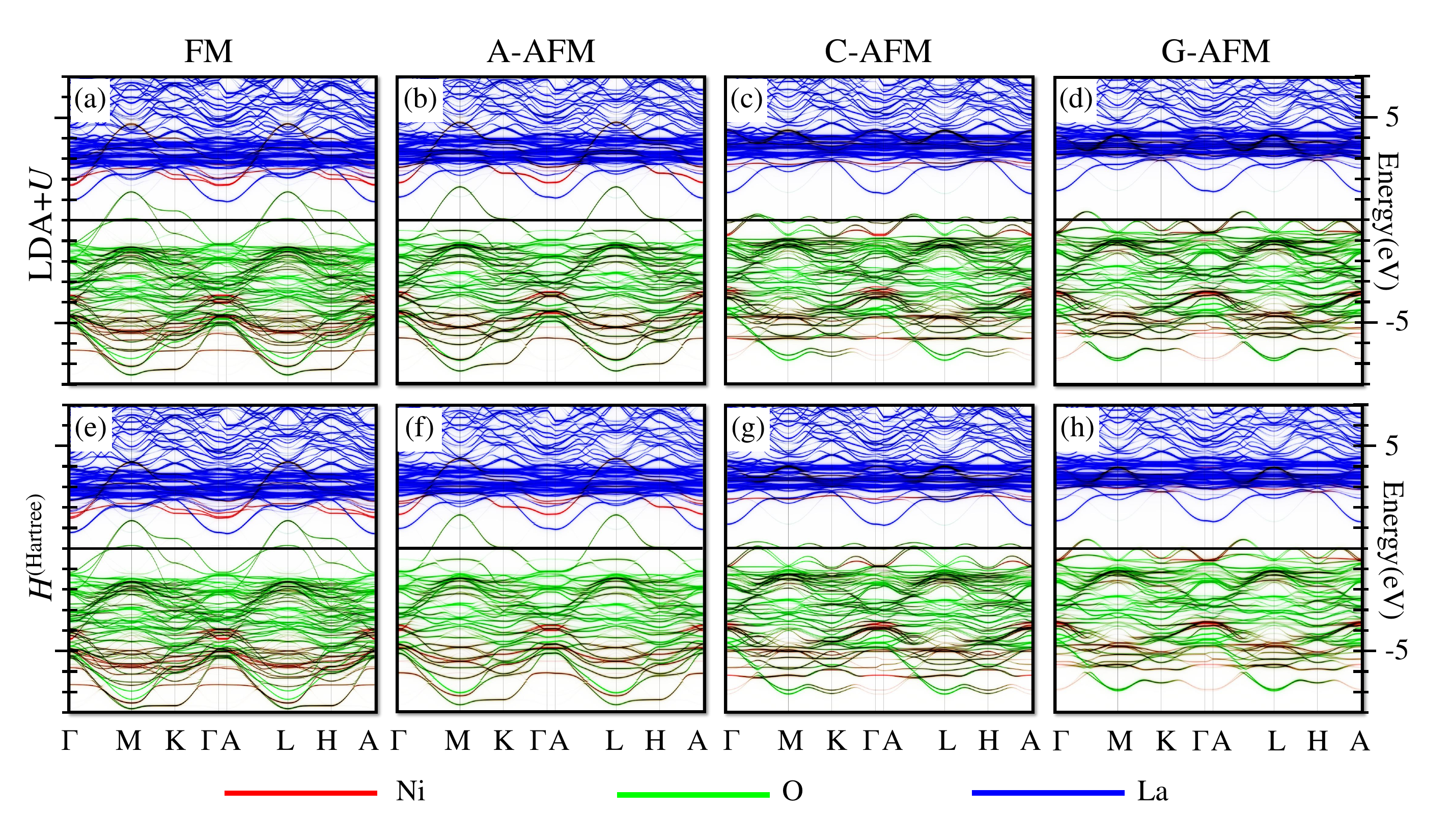}
	\end{center}
	\vspace{-0.8cm}
	\caption{
	Quality verification of $H^{^\mathbf{(Hartree)}}$ for the low-pressure phase, by reproducing the LDA+$U$ band structures under (a) ferromagnetic, (b) A-type anti-ferromagnetic, (c) C-type anti-ferromagnetic, and (d) G-type anti-ferromagnetic order, with its self-consistent Hartree-Fock results in (e)-(h) using a fixed set of parameter.
	}
	\label{figs1}
	\vspace{-0.5cm}
\end{figure}

\begin{figure}[h]
	\begin{center}
	\includegraphics[width=1\columnwidth]{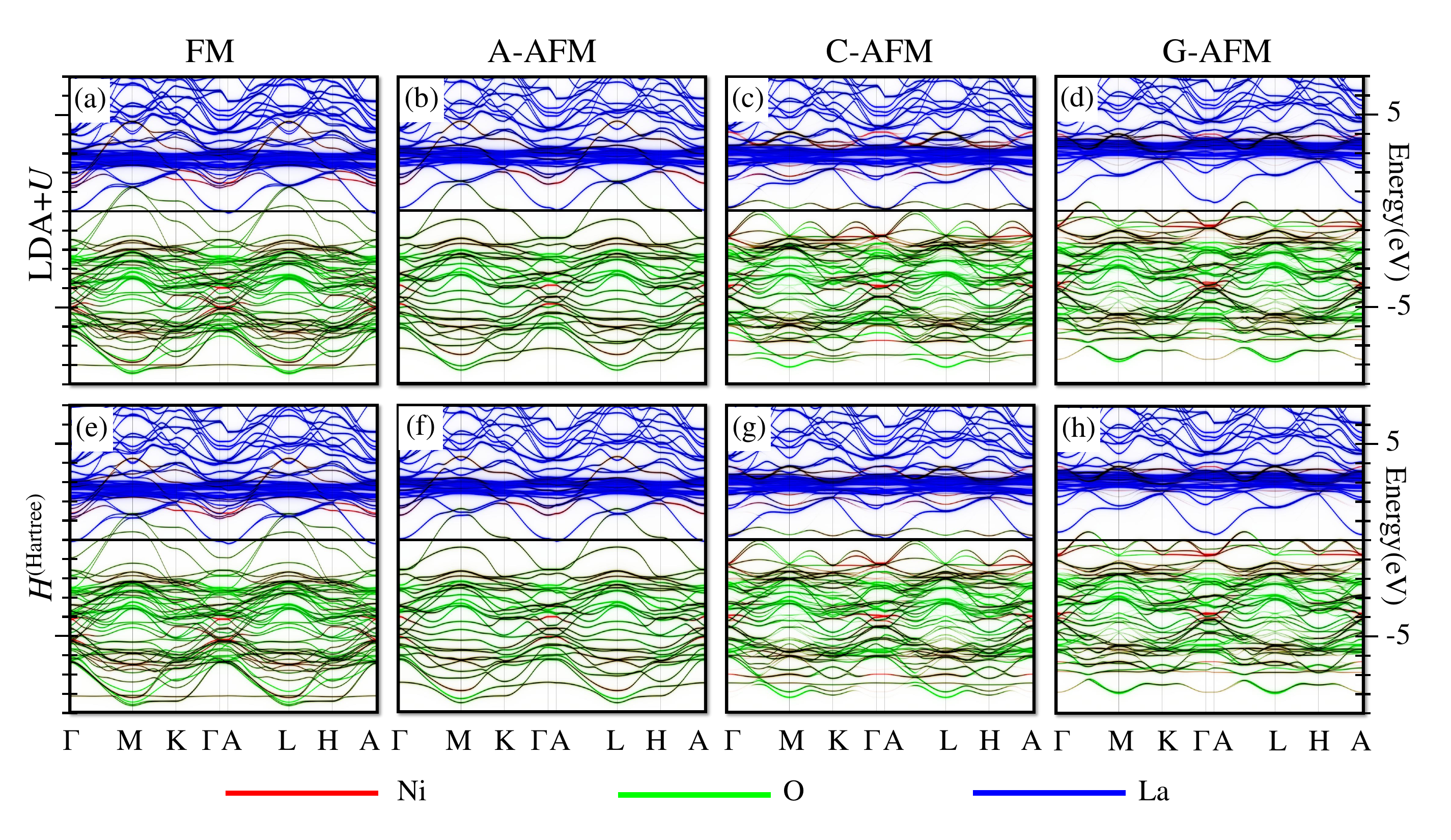}
	\end{center}
	\vspace{-0.8cm}
	\caption{
    Same as Fig.~\ref{figs1}, but for the high-pressure phase.
	}
	\label{figs2}
	\vspace{-0.1cm}
\end{figure}

We thus fix the parameters in Eq.~\ref{eq1} by requiring it to reproduce the DFT+$U$ band structure under \textit{various} magnetic structures (with a single set of parameters)~\cite{Wei2006,lang2021,Jiang2022}, via the same self-consistent Hartree-Fock approximation as employed in DFT+$U$~\cite{Anisimov1997}.
The procedure can be simplified if one assumes that the structure of $U_{m m^{\prime\prime} m^{\prime}m^{\prime\prime\prime}}$ follows the Slater integral~\cite{Slater,Liechtenstein1995} fixed by two parameters $U_{\text{eff}}$ and $J_{\text{eff}}$.
Furthermore, we find that the $t$ parameters are quite insensitive to the magnetic structure, so one typically can just take the spin-average of that obtained from Wannier functions.
The next section lists some of the leading parameters.

Figure~\ref{figs1} and \ref{figs2} compare the DFT+$U$ band structures and those from Hartree-Fock approximation of our extracted $H^\text{(Hartree)}$, under various common magnetic structures: ferromagnetic and A-type, C-type, G-type antiferromagnetic order.
One see that for the low-pressure phase, a single $H^\text{(Hartree)}$ with a fixed set of parameters is able to reproduce nicely all the DFT+$U$ band structure.
The same is true also for the high-pressure phase, with its own fixed set of parameters.
These comparison therefore confirms the quality of our obtained $H^\text{(Hartree)}$ and the corresponding parameters for both phases.

Even though the quality of $H^{\text{(Hartree)}}$ is gauged by comparing its approximate solutions against the DFT+$U$ results (under similar approximation), the utilization of $H^{\text{(Hartree)}}$ is obviously not limited to such approximate treatment, nor is it necessary for $H^{\text{(Hartree)}}$ to generate a magnetically ordered phase.
As we demonstrate in this work, such a SU(2) symmetric interacting Hamiltonian can allow many interesting physics to emerge at low energy.

\vspace{-0.2cm}
\subsection{7. Leading parameters for $H^{\text{(Hartree)}}$}

\begin{table}
    \caption{One-body kinetic parameters $t$ in $H^{\mathrm{(Hartree)}}$ in unit of eV for Ni-$d_{3z^2-r^2}$, $d_{x^2-y^2}$, in-plane O$^{\parallel}$-$p_x$, and apical O$^{\perp}$-$p_z$ orbitals, for the low-pressure and high-pressure phases.}
    \begin{ruledtabular}
        \begin{tabular}{lllll}
(low-pressure) & O$^{\perp}$-$p_z$  & Ni-$d_{3z^2-r^2}$ & O$^{\parallel}$-$p_x$ & Ni-$d_{x^2-y^2}$ \\ \hline
O$^{\perp}$-$p_z$     & -2.341             &\textbf{1.392}    &0.515                 & 0 \\ 
Ni-$d_{3z^2-r^2}$     & \textbf{1.392}    &-1.451             &0.772                  & 0 \\
O$^{\parallel}$-$p_x$ & 0.515             & 0.772             &-3.577                 & 1.481 \\
Ni-$d_{x^2-y^2}$      & 0                  & 0                 & 1.481                 & -1.158 \\ \hline\hline

(high-pressure) & O$^{\perp}$-$p_z$  & Ni-$d_{3z^2-r^2}$ & O$^{\parallel}$-$p_x$ & Ni-$d_{x^2-y^2}$ \\ \hline
O$^{\perp}$-$p_z$     & -4.405             &\textbf{2.179}    &0.513                 & 0 \\ 
Ni-$d_{3z^2-r^2}$     & \textbf{2.179}    &1.567             &0.905                  & 0 \\
O$^{\parallel}$-$p_x$ & 0.513             & 0.905             &-4.222                 & 1.623 \\
Ni-$d_{x^2-y^2}$      & 0                  & 0                 & 1.623                 & -1.393 \\
\end{tabular}
    \end{ruledtabular}
    \label{tabs1}
    \vspace{-0.5cm}
\end{table}

Using the above procedure, we find ($U_{\text{eff}}$, $J_{\text{eff}}$) $\sim$ (6.12, 1.12)eV in the low-pressure phase and ($U_{\text{eff}}$, $J_{\text{eff}}$) $\sim$ (5.99, 1.18)eV in the high-pressure phase.
Other leading parameters extracted from the procedure are given in  
Tab.~\ref{tabs1}, concerning the four most relevant orbitals, namely Ni-$d_{3z^2-r^2}$, $d_{x^2-y^2}$, O$^{\parallel}$-$p_x$, and O$^{\perp}$-$p_z$.
(Here O$^{\parallel}$-$p_x$ denotes the $p$-orbitals pointing toward the neighboring Ni atoms.)
The full $t_{ii^\prime mm ^\prime}$ parameters are available upon request.

\subsection{8. Obtaining low-energy effective Hamiltonian via numerical canonical transformation}

Low-energy effective theory is one of the most effective means to capture and understand key physics of a quantum system within a well-defined energy scale.
Conceptually, it can be rigorously constructed by integrating out the high-energy states in the path integral formulation.
Equivalently, it can also be derived by decoupling the remaining low-energy states from the high ones.
Specifically, one finds a unitary transformation\cite{chao1977kinetic,White2002}, $U[\{a_i,a^{\dagger}_i\}]$, of the second quantized basis, $a_i$, spanning the one-body space indexed by $i$,
\begin{equation}
    \tilde{a_i}=Ua_iU^\dagger,
    \label{unit_basis}
\end{equation}
such that the Hamiltonian has a `block diagonal' form, $\Tilde{H}$, in the new `many-body dressed' representation, $\tilde{a_i}$ (freeing the low-energy subspace from influence of the high-energy subspace),
\begin{equation}
    H[\{a_i,a^{\dagger}_i\}] = \Tilde{H}[\{\tilde{a}_i, \tilde{a}^\dagger_i\}].
    \label{eq2}
\end{equation}
Given
\begin{equation}
\begin{aligned}
    \Tilde{H}[\{\tilde{a}_i, \tilde{a}^\dagger_i\}] =& \Tilde{H}[\{Ua_iU^\dagger, Ua^\dagger_i U^\dagger\}]\\
    =& U \Tilde{H}[\{a_i,a^\dagger_i\}] U^\dagger.
    \label{eq3}
\end{aligned}
\end{equation}
the unitary transformation $U$ can thus be found by demanding
\begin{equation}
    \tilde{H}[\{a_i,a^\dagger_i\}] = U^\dagger H[\{a_i,a^\dagger_i\}] U,
    \label{eq4}
\end{equation}
to possess the desired block diagonal form.

Such block diagonalization can be performed numerically following a series of such unitary transformations until the unwanted off-diagonal block, $P^{\text{(L)}} H P^{\text{(H)}}$, is suppressed below some tolerance, with $P^{\text{(L)}}$ and $P^{\text{(H)}}\equiv 1-P^{\text{(L)}}$ being the projection operator for the low- and high-energy subspace.
Particularly, if the `rotational angle' of each transformation is small, each unitary transformation can be performed using just the low order terms in the expansion
\begin{equation}
    \tilde{H} = e^A H e^{-A}= H + [A,H] + \frac{1}{2!}[A,[A,H]] +\cdots,
\end{equation}
using a `small' anti-Hermitian operator $A$ that generates each unitary transformation $U^\dagger=e^A$.
Specifically, for each term $C$ in $P^{\text{(L)}} H P^{\text{(H)}}$, $A$ then includes a contribution $\alpha (C-C^{\dagger})$ with sign (and size) of $\alpha$ chosen properly to reduce the resulting unwanted block.
In some cases, special care should be taken when choosing $\alpha$ for each contribution of $A$ to ensure the proper symmetry of the representation~\cite{Hou_prepation}.

Upon reaching a satisfactory level of accuracy with negligible $P^{\text{(L)}} H P^{\text{(H)}}$, the low-energy effective Hamiltonian  can then be taken from the low-energy sector of $\Tilde{H}$, \begin{equation}
    H^\prime = P^{\text{(L)}} \Tilde{H} P^{\text{(L)}}.
\end{equation}
Obviously, the above procedure can be repeated to obtain even lower-energy scale Hamiltonians by identifying another set of leading high-energy states to `integrate out' and construct the corresponding projection operators.

Practically, we build a set of C++ codes to perform the above procedure algebraically, by manipulating second-quantized operators numerically.
This includes, for example, evaluation of commutation and perform normal ordering required for the procedure.

A significant reduction of computation expense can be achieved if one restricts the Hilbert space of interest \textit{a priori}.
As an important example, most physical Hamiltonians conserve particle number, $N$, so do the corresponding $A$ and $U$.
One can therefore apply a projection operator $P_N$ of the subspace with fixed particle number $N$ to $H \rightarrow P_N H P_N$, or in any stage of the procedure.
$P_N$ can be practically implemented as $P_N=P_{\leq N}~P_{\geq N}$, where $P_{\geq N}$ is the $N$-body representation of operator 1 in second-quantized notation that ensures at least $N$ particles present in the system.
$P_{\leq N}$ on the other hand corresponds to truncation of all terms higher than $N$-bodies in any stage of the computation.
In many cases, we find that applying it to $P^{(L)}\rightarrow P_NP^{(L)}$, if possible, is the easiest and most efficient.

In this particular study, three sets of high-energy states are integrated out at different energy scale.
At the Hartree scale, they are the states containing charged Ni ion with $d^9$ configuration.

The corresponding projection operator is therefore 
\begin{equation}
\begin{split}
    P^{\text{(L)}}&=(1-X^{\dagger}_{\uparrow}X^{\dagger}_{\downarrow}X_{\downarrow}X_{\uparrow})^\text{(1)}(1-Z^{\dagger}_{\uparrow}Z^{\dagger}_{\downarrow}Z_{\downarrow}Z_{\uparrow})^\text{(1)}\\& \cross(1-X^{\dagger}_{\uparrow}X^{\dagger}_{\downarrow}X_{\downarrow}X_{\uparrow})^\text{(2)}(1-Z^{\dagger}_{\uparrow}Z^{\dagger}_{\downarrow}Z_{\downarrow}Z_{\uparrow}  )^\text{(2)},
\end{split}
\end{equation}
where $X^\dagger_\uparrow$ and $Z^\dagger_\uparrow$ denotes creation of a spin-up electron in the $d_{x^2-y^2}$ and $d_{3z^2-r^2}$ orbital, respectively, of Ni in the upper (1) and lower (2) layer of the Ni-O-Ni bi-layer.
Further absorbing the particle number projection $P_N$ for $N=6$ to $P^{(L)}$ gives
\begin{equation}
\begin{split}
    P_NP^{\text{(L)}}&=p^{\dagger}_{\uparrow}p^{\dagger}_{\downarrow}p_{\downarrow}p_{\uparrow} (X^{\dagger}_{\uparrow}X_{\uparrow}+X^{\dagger}_{\downarrow}X_{\downarrow})^\text{(1)}(Z^{\dagger}_{\uparrow}Z_{\uparrow}+Z^{\dagger}_{\downarrow}Z_{\downarrow})^\text{(1)}\\& \cross(X^{\dagger}_{\uparrow}X_{\uparrow}+X^{\dagger}_{\downarrow}X_{\downarrow})^\text{(2)}(Z^{\dagger}_{\uparrow}Z_{\uparrow}+Z^{\dagger}_{\downarrow}Z_{\downarrow})^\text{(2)}
    P^{\text{(L)}},
\end{split}
\end{equation}
where $p^\dagger_\uparrow$ denotes creation of a spin-up electron in the $p_z$ orbital of the apical O$^\perp$ between the Ni atoms.
As discussed in the main text, $P_NP^{(L)}$ above corresponds to a low-energy subspace with double occupation of $p_z$ and single occupation for the rest.
Applying the above procedure then gives $H^\text{(eV)}=P_NP^{(\text{L})}\tilde{H}P^{(\text{L})}$ in Eq.(1).

At eV-scale, for the Hund's coupling $\tilde{J}_\text{H}$-dominant low-pressure phase, the high-energy states are those containing a spin singlet $W^\dagger\ket{0}$ among the $d_{3z^2-r^2}$ and $d_{x^2-y^2}$ orbitals within the same Ni ion, denoted by singlet creation operator,
 \begin{equation}
     W^\dagger=\frac{1}{\sqrt{2}}(X^{\dagger}_{\uparrow} Z^{\dagger}_{\downarrow}-X^{\dagger}_{\downarrow} Z^{\dagger}_{\uparrow}).
 \end{equation}
Applying the above procedure with the projection operator,
\begin{equation}
     P^{\text{(L)}}= (1-W^\dagger W)^{(1)}(1-W^\dagger W)^{(2)},
\end{equation}
results in $H^{\text{(sub-eV)}}$ in Eq.(2) with spin-1 effective Ni$^{2+}$ ionic spin  $\mathbf{S}_{\mathrm{eff}}$.

In contrast, for the high-pressure phase dominated by inter-layer anti-ferromagnetic super-exchange $J_{ZZ}$, the high-energy states are those containing spin triple among the $d_{x^2-y^2}$ orbitals of the Ni ions across the Ni-O-Ni bi-layer.
Expressed via the creation operator of the remaining \textit{inter}-layer singlet,
\begin{equation}
     V^\dagger= 
     \frac{1}{\sqrt{2}}[Z^{\dagger(1)}_{\uparrow}Z^{\dagger(2)}_{\downarrow}-Z^{\dagger(1)}_{\downarrow}Z^{\dagger(2)}_{\uparrow}],
\end{equation} 
the projection operator for the low-energy subspace is simply,
\begin{equation}
     P^{\text{(L)}}=V^\dagger V.
\end{equation}
The above procedure then gives $H^{\text{(sub-eV)}}$ in Eq.(2) with a fractionalized  spin-1/2 effective ionic $\mathbf{S}_{\mathrm{eff}}$ for each Ni$^{2+}$ ion.

\subsection{9. Alternative implementation for derivation of effective Hamiltonian}
\begin{figure}
\centering
\includegraphics[width=1\columnwidth]{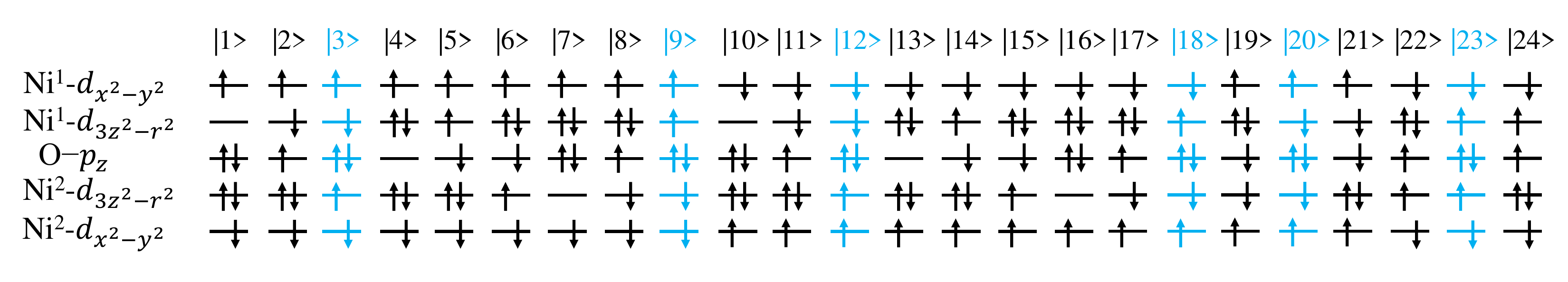}
\caption{Illustration of local many-body states of Ni-O-Ni bi-layer component in the subspace with zero $z$-component of total spin. The states in blue are the low-energy $|d^{8}p^{6}d^{8}>$ states without Ni ions in the $d^9$ configuration.
}
\vspace{-0.2cm}
\label{Sz0_states_total}
\end{figure}

\begin{figure}
\centering
\includegraphics[width=1\columnwidth]{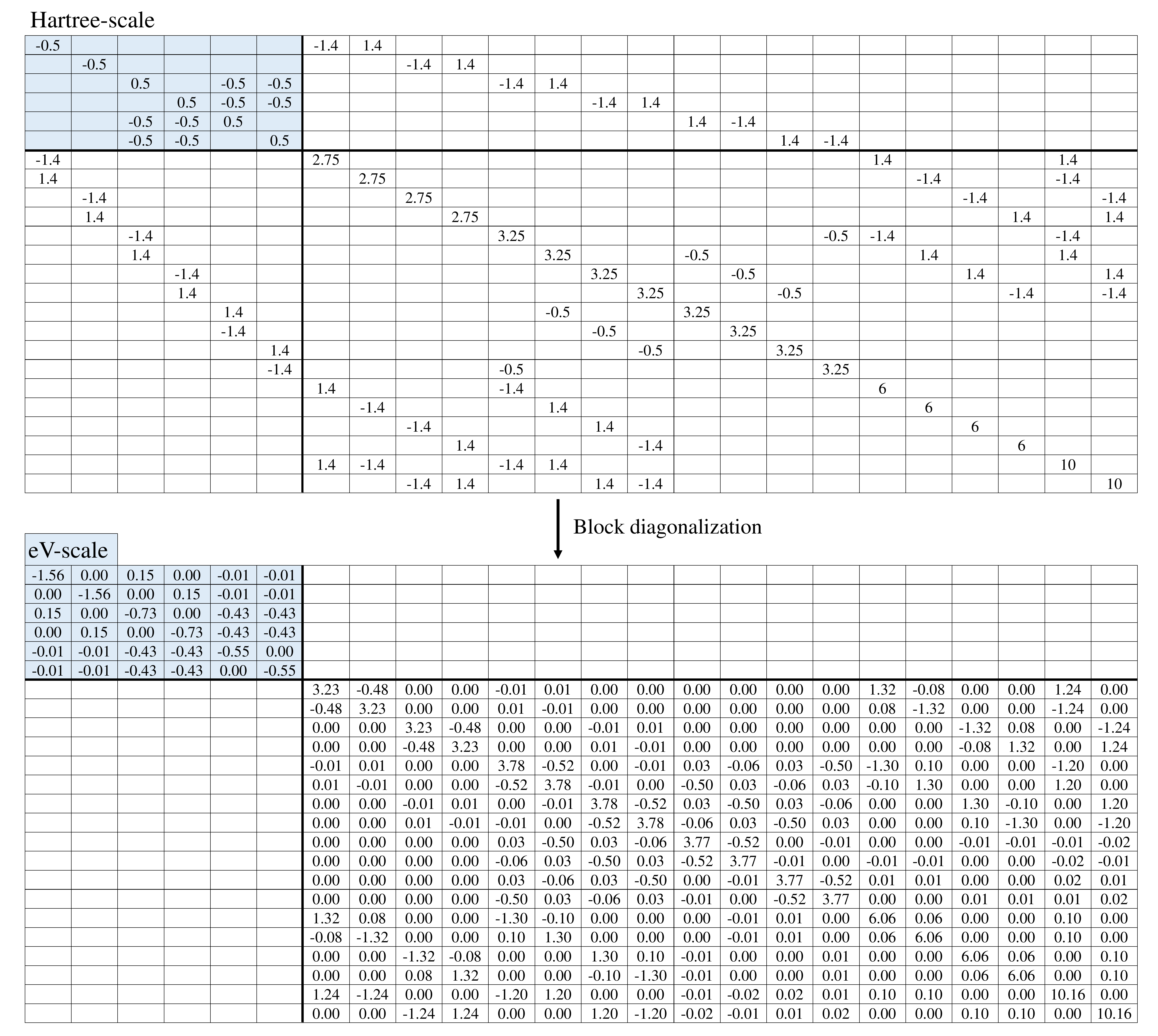}
\caption{An example of deriving $H^{\text{(eV)}}$ from $H^{\text{(Hartree)}}$ (in unit of eV) via block diagonalization in the subspace with zero $z$-component of total spin.
The resulting upper left corner block in blue represents $H^{\text{(eV)}}$ in the low-energy subspace.
Empty cells correspond to those with identical zero values (up to machine accuracy), while `0.00' represents values smaller than 0.01 in size.
}
\vspace{-0.4cm}
\label{matrix_eV}
\end{figure}

\begin{figure}
\centering
\includegraphics[width=1\columnwidth]{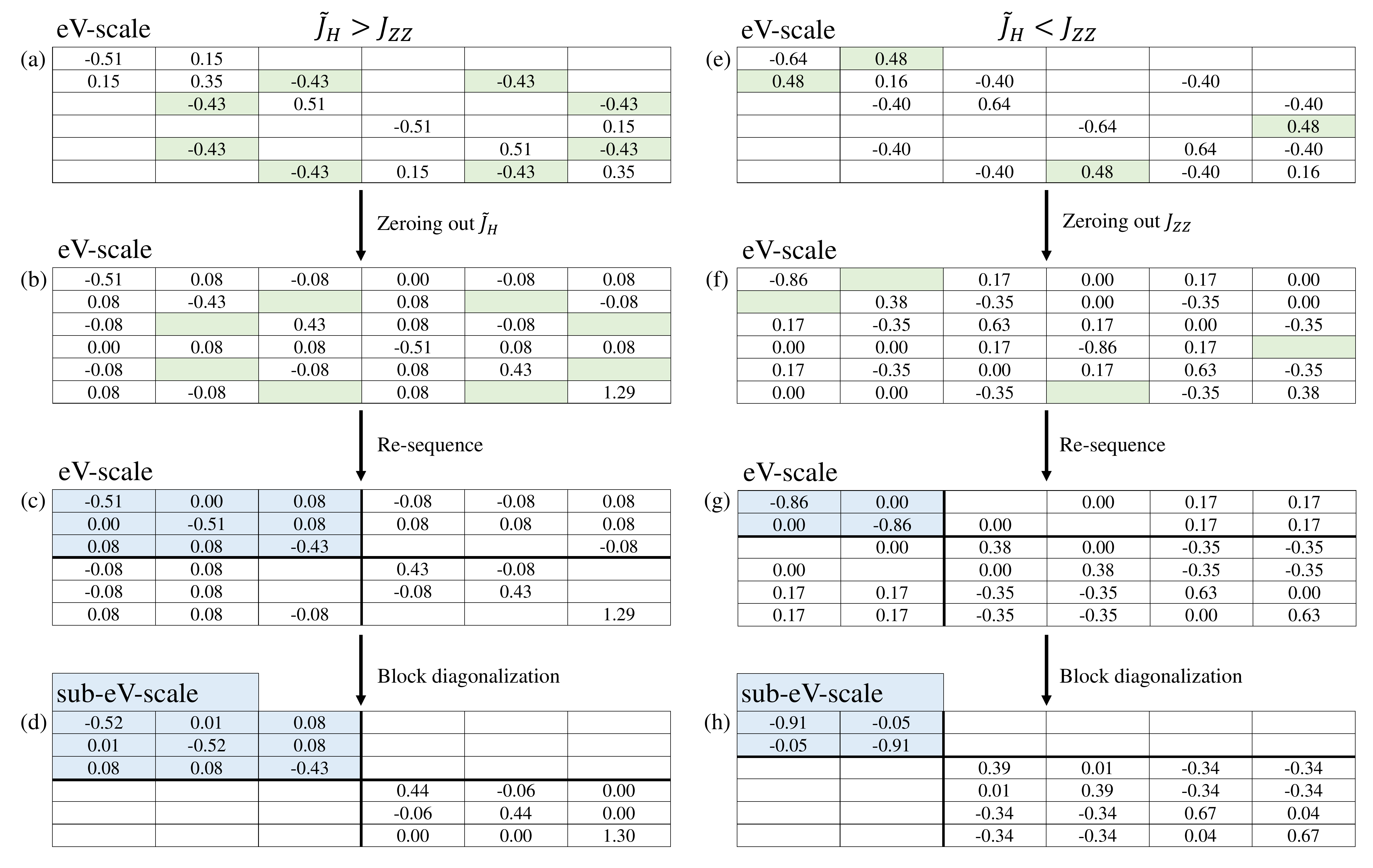}
\caption{Examples of deriving $H^{\text{(sub-eV)}}$ from $H^{\text{(eV)}}$ (in unit of eV) in the subspace with zero $z$-component of total spin.
(a) Identify the leading physics ($\tilde{J}_\text{H}$) and the corresponding off-diagonal terms in green.
(b) Zero out these terms via unitary transformation to identify the low-energy subspace (with spin-1 Ni ions).
(c) Re-sequence for an easier visualization.
(d) Decouple the low-energy sector (in blue) from the rest via block diagonalization. 
The resulting upper left block gives the low-energy effective $H^{\text{(sub-eV)}}$.
Empty cells correspond to those with identical zero values (up to machine accuracy), while `0.00' represents values smaller than 0.01 in size.
The same procedure is applied to the other case $\tilde{J}_{\mathrm{H}}<J_{ZZ}$ (e)-(h).
Notice that by (b) and (f) the low-energy subspace already reflects the fractionalization of the Ni$^{2+}$ ionic spin from 1 to 1/2.
}
\vspace{-0.4cm}
\label{matrix_sub_eV}
\end{figure}

Instead of second quantized operators and their non-commuting algebra, in this section we demonstrate an example of deriving low-energy effective Hamiltonian using an alternative implementation of the above general idea in the standard \textit{first-quantized} representation via the vector space of the many-body states.

Figure.~\ref{Sz0_states_total} lists the many-body states that spans a subspace with zero $z$-component of the total spin that are decoupled from the rest of the many-body states.
First, we identify the low-energy states as those without Ni-$d^9$ configuration, and label them in blue.
For clarity, we re-sequence these states to group together the low-energy $\{|9>, |12>, |3>, |18>, |20>, |23>\}$ and high-energy states $\{|5>, |8>, |11>, |15>, |2>, |6>, |14>, |17>, |19>, |21>, |22>, |24>, |1>, |7>, |10>, |16>, |4>, |13>\}$, when representing $H^{\text{(Hartree)}}$ as the upper matrix in Fig.~\ref{matrix_eV}.
$H^{\text{(Hartree)}}$ is then numerically block-diagonalized into the lower matrix in Fig.~\ref{matrix_eV}, via a series of symmetric unitary transformation.
The low-energy blue block of that matraix therefore represents the low-energy effective Hamiltonian $H^{\text{(eV)}}$, which can be mapped to a second-quantized expression of $H^{\text{(eV)}}$ in Eq.(1).

The same procedure can be repeated for the derivation of even lower-energy Hamiltonians from $H^\text{(eV)}$ shown in Fig.~\ref{matrix_sub_eV}(a).
For the low-pressure phase, the next dominant physics is the ferromagnetic renormalized Hund's coupling $\tilde{J}_\text{H}\sim0.86$ within each Ni ion, whose effects are \textit{not} diagonal in our starting representation [observe the off-diagonal elements $\tilde{J}_\text{H}/2 \sim 0.43$ in the matrix.]
One therefore cannot identify the low-energy and high-energy Hilbert space based solely on the diagonal elements.
We therefore first perform a series of symmetric unitary transformations to zero out \textit{only} those off-diagonal elements associated with contributions of $\tilde{J}_{\text{H}}$.
This effectively finds a new set of basis states to represent $H^{(\text{eV})}$ shown in Fig.~\ref{matrix_sub_eV}(b), whose diagonal elements now distinguish the high-energy states (those with spin-0 Ni ion) from the low-energy ones.
Fig.~\ref{matrix_sub_eV}(c) gives an easier visualization upon re-sequence the basis that sorts the diagonal elements.
We can now proceed to block-diagonalize matrix in (c) into the one in (d) and extract the corresponding $H^{(\text{sub-eV})}$ from the blue low-energy section as before.

As a remark, since we employ a complete decoupling between the remaining low-energy subspace from the rest in this study, the resulting eigenstates of the low-energy sector is guarenteed to be a true eigenstates of the original high-energy Hamiltonian.
The completeness of the remaining low-energy states can be easily verified against those from the high-energy Hamiltonian.

Following the same procedure described above, Fig.~\ref{matrix_sub_eV}(e)-(h) also gives an example for the high-pressure phase, where the leading physics at the eV scale is the inter-layer super-exchange $J_{ZZ}$.
Notice that in (f) or (g), the low-energy subspace already is of different size from that of the low-pressure phase in (b) and (c), reflecting the fractionalization of Ni$^{2+}$ effective spin $\textit{S}_{\text{eff}}$ from 1 to 1/2.

\subsection{10. Parameters in deriving $H^{\text{(eV)}}$ from $H^{\text{(Hartree)}}$ }

In deriving the local $H^{\text{(eV)}}$ from $H^{\text{(Hartree)}}$, only a few effective parameters are needed, most of which can be directly obtained from $H^{\text{(Hartree)}}$.
The kinetic energy $t_{pZ}^{\perp}$ is $\sim1.4$eV and $\sim2.2$eV for the low-pressure and high-pressure, respectively.
The $\ket{d^8p^6d^8}$ to $\ket{d^9p^5d^8}$ charge transfer energy $E_\mathrm{CT}$ is $\sim 3$eV.
The Coulomb interaction $U_d$ and Hund's coupling $J_{\mathrm{H}}$ among Ni-$d$ orbitals are $\sim6$eV, and $\sim1$eV, respectively.
Note that due to nearly full occupation of O-orbitals, the LDA+$U$ calculation is insensitive to the value of intra-atomic Coulomb repulsion $U_{p}$ of O-orbitals.
$U_p$ is therefore absorbed as local site energy of O orbitals in $H^{\text{(Hartree)}}$, and later estimated to be $\sim4$eV following the literature~\cite{Ogata_2008}.
We checked that our qualitative conclusions are insensitive to 1eV variation of $U_p$.

\subsection{11. Rapid growth of $J_{ZZ}$ upon increasing $t^\perp_{pZ}$ through applied pressure}

\begin{figure}[h!]
\centering
\includegraphics[width=0.85\columnwidth]{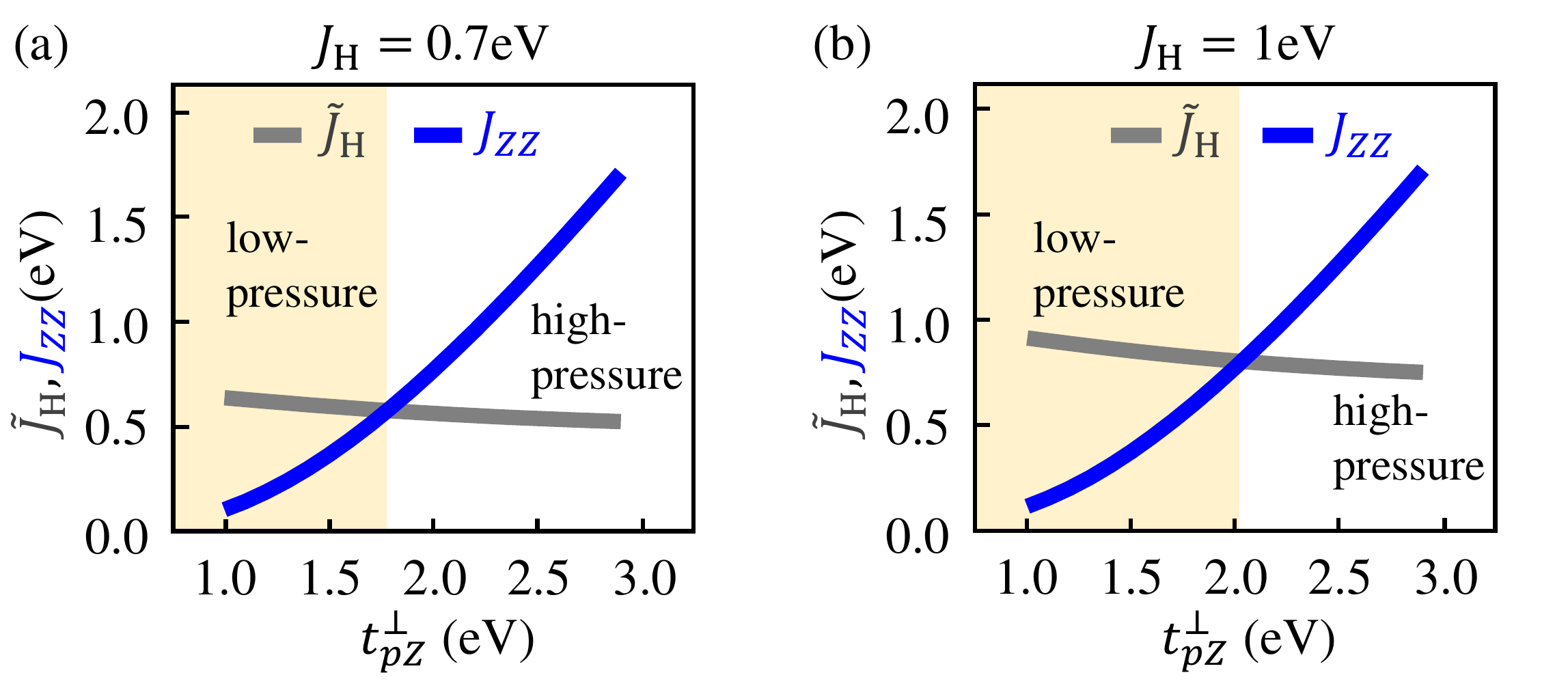}
\caption{Rapid growth of inter-layer super-exchange $J_{ZZ}$ at eV-scale due to enhanced $t^{\perp}_{pZ}$ under higher pressure.
In contrast, the intra-atomic Hund's coupling $J_\mathrm{H}$ [0.7eV in (a) and 1.0eV in (b)] is weakly screened.
Distinct from the low-pressure phase, the inter-layer super-exchange dominates in the high-pressure phase, leading to fractionalization of the Ni$^{2+}$ ionic spin at sub-eV scale.
Note that inclusion of additional fluctuation from the itinerant carriers to the spin in $d_{x^2-y^2}$ orbital will weaken the effectiveness of $J_\mathrm{H}$ and in turn further strength the high-pressure phase.
}
\vspace{-0.4cm}
\label{S3}
\end{figure}

Microscopically, the qualitative change of local physics shown in this work results from the enhancement of $J_{ZZ}$ at the eV scale due to strengthened $t^{\perp}_{pZ}$ at high-pressure.
Figure.\ref{S3} shows that this enhancement is rather rapid due to the high-order nature of its emergence.
(In perturbable regime, it is proportional to $t_{pZ}^{\perp 4}$.)
This trend should be further enhanced upon inclusion of additional fluctuation from the itinerant carriers to the spin in $d_{x^2-y^2}$ orbital, as it can only weaken the effectiveness of $J_\mathrm{H}$ and in turn further strength the high-pressure phase, similar to the trend from panel (b) toward panel (a).
The observed first-order like transition, associated with the switching of leading physics from $\tilde{J}_{\mathrm{H}}$ to $J_{ZZ}$ at eV scale, is therefore inescapable (requires no fine-tuning.)

\bibliography{Main_v2.bib}

\begin{thebibliography}{83}
\bibitem [{not()}]{note_hole_impact}%
  \BibitemOpen
  \href@noop {} {}\bibinfo {note} {Introduction of the remaining ligand hole
  will not qualitatively affect our study, since for the local Ni-O-Ni
  component, its main effect is to fluctuate the $d_{x^2-y^2}$ orbital,
  effectively reducing the Hund's coupling and thus enhancing the main physics
  found at high-pressure.}\BibitemShut {Stop}%
\end{thebibliography}
\end{document}